\documentstyle[prl,aps,epsf]{revtex}
\draft
\begin{document}
\twocolumn[\hsize\textwidth\columnwidth\hsize\csname @twocolumnfalse\endcsname

\title{Novel Colloidal Crystalline States on Two Dimensional Periodic
Substrates} 
\author{C. Reichhardt and C.J. Olson} 
\address{ 
Center for Nonlinear Studies, Theoretical, and Applied Physics Divisions, 
Los Alamos National Laboratory, Los Alamos, NM 87545}

\date{\today}
\maketitle
\begin{abstract}
We show using numerical simulations that a rich variety of novel colloidal
crystalline states are realized on square and triangular two dimensional
periodic substrates which can  be
experimentally created using crossed laser arrays. 
When there are more colloids than potential substrate minima, 
multiple colloids are trapped at each substrate minima and act
as a single 
particle with a rotational degree of freedom, giving
rise to a new type of orientational order.  
We call these states colloidal molecular crystals.  
A two-step melting can also  occur in which
individual colloidal molecules initially rotate, 
destroying the overall orientational order,
followed by the onset of inter-well colloidal hopping. 
\end{abstract}
\vspace{-0.1in}
\vspace{-0.3in}

\vskip2pc]
\narrowtext
Colloidal particles   
are an ideal system for studying 2D ordering and melting 
as the individual particle positions and dynamics 
can be directly visualized. Colloidal melting has been 
extensively studied for systems with smooth substrates,
where evidence for defect-mediated
melting transitions has been found \cite{Grier,Nm,Lowen}.   
Another avenue of study has been the examination of 
colloidal crystallization and melting in the 
presence of a 1D periodic
substrate, which are typically created with laser 
arrays \cite{Bechinger,Leiderer,Sood,Frey,Lin,Rochon}.    
These studies have revealed a novel laser-induced freezing, as well 
as a reentrant laser induced melting
as a function of substrate strength \cite{Leiderer,Frey}. 
A far less studied case is colloidal crystallization on 
2D periodic substrates. 
Here numerous commensuration effects can be
expected when the periodicity of the colloidal
lattice matches with the substrate lattice constant. 
Several recent experimental 
studies have demonstrated that it is possible to 
create a 2D substrate for colloids 
using optical tweezer arrays \cite{Korda}, 
templating\cite{Lin,Ling}, and 
2D crossed laser arrays \cite{Bech}.
Almost nothing is known, however, about what 
type of colloidal crystalline states could occur and 
what  types of melting can occur in this
system. 
Although colloids interacting with 1D substrates have been
investigated theoretically \cite{Sood,Frey} and 
numerically \cite{Sood}, the case of a 2D 
substrate has not.     
The study of
colloidal crystals with 2D substrates 
may also lend insight into other  
condensed matter systems which can be described as elastic
particles on 2D substrates.  Examples of such systems include 
atoms on atomic surfaces \cite{Proko} 
and vortices in
superconductors with periodic pinning arrays \cite{Reichhardt}. 
It should also be possible to realize entirely new states 
in the colloidal system, since when there are more colloids than substrate
minima, each minima will contain multiple colloids which can have
their own internal degrees of freedom.  
In addition, the control of different types of  
colloidal crystal geometries can be of technological importance  
for the construction of certain devices, such as photonic
band-gap materials. 

In this work we examine colloidal ordering and melting on 2D periodic
square and triangular substrates using Langevin simulations. 
We find that a 
rich variety of novel colloidal crystalline states can be achieved.
We focus on the case of filling fractions at which there are an 
integer number of colloids per substrate minima.
When there are more colloids than potential substrate minima, 
multiple colloids are trapped at each minima. 
The colloids within each of these minima can act as a single particle 
with internal degrees of rotational freedom, forming trimer and dimer
states that have an additional 
long-range orientational order with respect to neighboring
minima.
These states are similar to molecular crystals; hence,  
we term them colloidal molecular crystals or CMC's. 
These crystalline states also exhibit a novel multi-stage melting where, 
for low $T$, the colloidal molecules possess both orientational 
and translational order. 
At higher $T$, the orientational order is lost as the 
colloidal molecules in each minima began to rotate, but 
individual colloidal diffusion does not occur 
and the colloids are still confined in each substrate minima. 
For large enough temperature, individual colloidal
diffusion occurs. We map out the phase diagram for temperature vs substrate
strength and find that two stage melting only occurs for sufficiently strong
substrates, 
and that the transition temperature from the solid to the disordered
solid in the presence of a substrate is lower than the same transition
temperature for a sample without a substrate.
In addition, for fixed 
temperature and increasing substrate strength, 
we observe a transition from the ordered solid phase
to a partially ordered solid 
phase, similar to the reentrant melting seen 
for colloids interacting with 1D periodic substrates. 

We simulate a 2D system 
of $N_{c}$ colloids
with periodic boundary conditions in the $x$ and
$y$ directions, using 
Langevin dynamics as employed in previous colloidal simulations  
\cite{Nm,Peeters}. The equation of motion for 
a colloid $i$ is:
\begin{equation}
\frac{d {\bf r}_{i}}{dt} = {\bf f}_{ij} + {\bf f}_{s} + {\bf f}_{T}
\end{equation}
Here ${\bf f}_{ij} = -\sum_{j \neq i}^{N_{c}}\nabla_i V(r_ij)$ 
is the interaction force from the other colloids. 
The colloid-colloid interaction 
between colloids $i$ and $j$ is a Yukawa or screened Coulomb
potential, 
$V(r_{ij}) = (Q^2/|{\bf r}_{i} - {\bf r}_{j}|)
\exp(-\kappa|{\bf r}_{i} - {\bf r}_{j}|)$, where

\begin{figure}
\center{
\epsfxsize=3.5in
\epsfbox{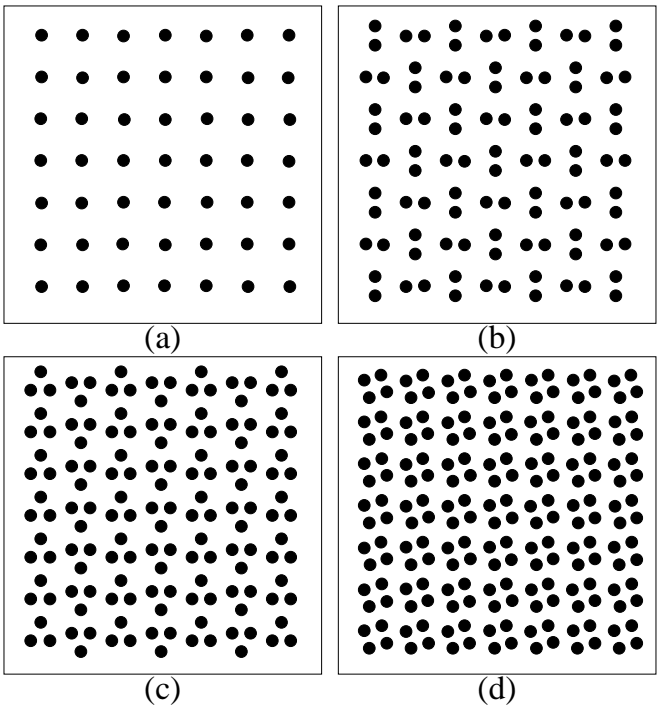}}
\caption{
The colloid configurations
(black dots) at $T = 0.0$ for
a square 2D periodic substrate with $A = 2.5$, for 
different densities of colloids with a fixed substrate period. 
(a) One colloid per potential minima with the colloids
forming a square commensurate lattice.  
(b) Two colloids per minima form
a colloidal dimer state, with each dimer perpendicular
to neighboring dimers.  
(c) Three colloids per minima
form a trimer state with orientational ordering. 
(d) Four colloids per minima produce an aligned quadrimer state.
}
\end{figure}

\hspace{-13pt}
$Q$ is the charge of the 
particles, $1/\kappa$ is the screening length, and ${\bf r}_{i(j)}$ is the 
position of
particle $i$ ($j$).
The length of the system is measured in units of the lattice constant 
$a_{0}$ and we take the screening length
$1/\kappa = a_{0}/2$.
For the force from the 2D substrate, ${\bf f}_{S}$,
we consider both square and triangular substrates with 
strength $A$, period fixed at $a_{0}$, and $N_{m}$ minima.
For square
substrates, ${\bf f}_{s} = A\sin(2\pi x/a_{0}){\hat {\bf x}} + 
A\sin(2\pi y/a_{0}){\hat {\bf y}}$, and for triangular substrates,
${\bf f}_{s}=\sum_{i=1}^{3}A\sin(2\pi p_{i}/a_{0})
[\cos(\theta_{i}){\hat {\bf x}}-\sin(\theta_{i}){\hat {\bf y}}]$,
where $p_{i}=x\cos(\theta_{i})-y\sin(\theta_{i})+a_{0}/2$,
$\theta_{1}=\pi/6$, $\theta_{2}=\pi/2$, and $\theta_{3}=5\pi/6$.
The thermal force ${\bf f}_{T}$ is a randomly fluctuating force 
from random kicks.    
We start the system at a temperature where all the colloids are diffusing
rapidly and gradually cool to $T = 0.0$. 
We define the melting temperature at zero substrate strength as $T_{m}^{0}$.
In this model, we do not take
into account hydrodynamic effects or possible long-range attractions between
colloids. 
We have conducted a wide variety of simulations for different system sizes,
$a_{0}$ values, and screening lengths $1/\kappa$,
and find that all qualitative results 
are robust. 

In Fig.~1 we show the colloidal positions for a
system with a square substrate. In Fig.~1(a) the colloidal periodicity 
matches one to one with the substrate periodicity, $N_{c}=N_{m}$,
so that each colloid 
is located at the center of 

\begin{figure}
\center{
\epsfxsize=3.5in
\epsfbox{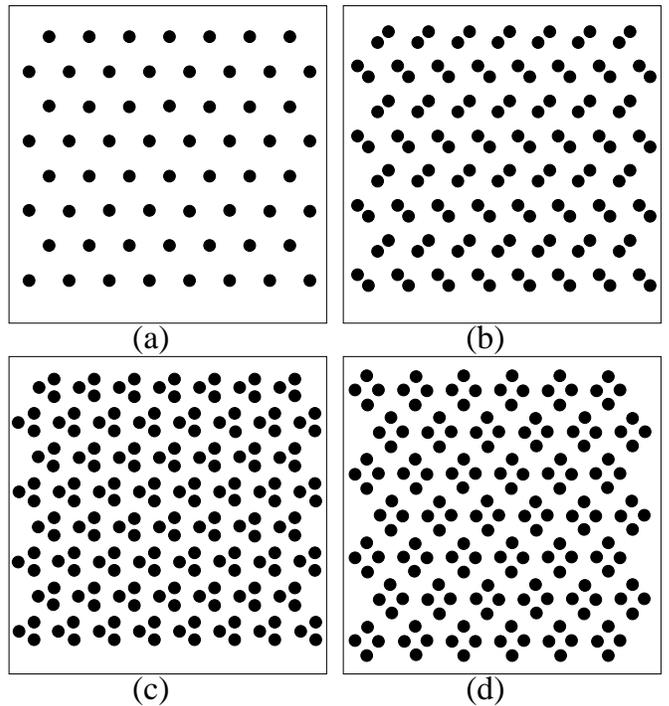}}
\caption{
The colloidal configuration
(black dots) at $T = 0.0$ after annealing a triangular
2D periodic substrate with $A = 2.5$ for an increasing density of
colloids and a fixed substrate period. (a) One colloid per substrate 
minima forms a commensurate triangular lattice. 
(b) Two colloids per minima form
an orientationally ordered dimer state,  
with the same dimer orientation in every other row. 
(c) Three colloids per minima form an orientationally
ordered trimer state.
(d)Four colloids per minima form a pattern of aligned diamonds.
}
\end{figure}

\hspace{-13pt}
the potential minima and a square colloidal crystal 
forms. In Fig.~1(b), for twice as many
colloids $N_{c}=2N_{m}$, each 
minima now captures two colloids.  Neither 
colloid is located at the exact minima; instead, they
are offset from it due to the colloid-colloid repulsion. The states inside
the minima can be regarded as a colloidal dimer with a rotational degree
of freedom.  
Fig.~1(b) also shows
that there are {\it two types of ordering} occurring: 
one arising from the square
substrate, and the other due to the 
specific rotational orientation of the colloidal dimers,
with neighboring dimers perpendicular to one another.       
The orientational ordering of the dimers is
due to the colloidal repulsion,
and allows the distance between the colloids to be maximized under the 
constraint of the square substrate. 
If the colloid interaction range is short,
such that one dimer pair does not interact with the neighboring pair,
the orientational dimer ordering seen in Fig.~1(b) 
does not occur. 
In Fig.~1(c), the colloidal configuration for $N_{c}=3N_{m}$
again shows two types 
of ordering, with each minima capturing three colloids
that form a trimer state. 
The trimers have a specific orientational order in which trimers in
adjacent columns of minima are rotated by 60$^\circ$, 
due to the repulsion of neighboring trimers. 
In 

\begin{figure}
\center{
\epsfxsize=3.5in
\epsfbox{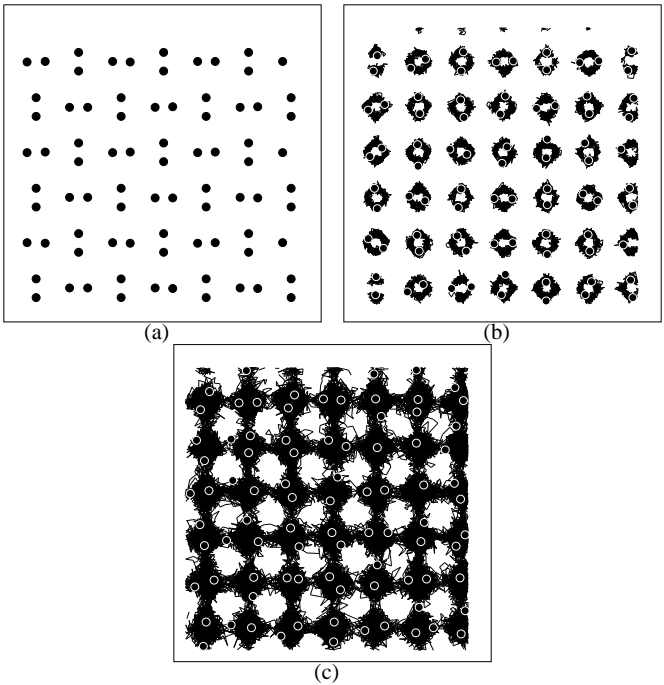}}
\caption{
The colloidal positions (black dots) and trajectories (lines) over
fixed time intervals at different temperatures 
for the
case of the dimer state on a square substrate shown in Fig.~1(b).
(a) $T/T^{0}_{m} = 0.25$, where
$T_{m}^{0}$ is the melting temperature at zero substrate
strength.  The system is in the
ordered solid phase and only small vibrations of the colloids around their
equilibrium positions occur. (b) $T/T_{m}^{0} = 1.5$.  The dimers
rotate within the substrate minima, destroying the orientational order
of the dimers, but the colloids remain trapped inside each minima
so the system is still in 
a solid phase. (c) $T/T_{m}^{0} = 4.0$.  Individual colloidal diffusion 
occurs throughout the sample and the system is in the liquid phase.  
}
\end{figure}

\hspace{-13pt}
Fig.~1(d), $N_{c}=4N_{m}$ also has
an ordered 
superlattice 
structure with orientational ordering. The colloidal 
quadrimer states are all aligned in the 
same 
direction.
For increasing colloidal densities $N_{c}=nN_{m}$
one can expect even more orderings
with each potential minima capturing $n$ colloids. 
Since the colloids in each
minima act as a single object 
with an internal degree of freedom, similar to a molecular dimer or trimer,
and since these states are confined in a lattice,  
we call these states colloidal molecular crystals (CMC's). 

In Fig.~2 we illustrate the CMC's on a triangular substrate.   
For one colloid per
minima, $N_{c}=N_{m}$, shown in Fig.~2(a), the colloids form a 
commensurate triangular lattice with 
each colloid located at the center of the potential minima.
For $N_{c}=2N_{m}$, a dimer state occurs
in each minima as seen in Fig.~2(b).
As in the square pinning case, 
an additional orientational ordering of the dimers occurs. 
The dimers in each row have the same orientation, which is rotated
45$^\circ$ with respect to the adjacent rows.
For $N_{c}=3N_{m}$, illustrated in Fig.~2(c), 
a timer state forms and 
an orientational ordering occurs in which,
unlike the square case [Fig.~1(c)], 
all the trimers have the {\it same} orientation. In Fig.~2(d) a
superlattice state also occurs with the colloid molecules forming a diamond
substructure with a triangular superstructure. 

We find that it is also possible to
create partially ordered and disordered 
states at fractional filling fractions, such as 
1.5 colloids per minima, $N_{c}=1.5N_{m}$. 
In this case, for the 
square substrates every other
minima forms a colloidal dimer state; however, the dimers do not have
long range orientational order. Similarly, for the triangular 
substrates it is not possible to arrange a  
state such that minima with only one colloid will be between
every other 
dimer, so again dimer ordering is not observed
at fractional filling. 

In Fig.~3 we show a representative
example of the two stage melting of the CMC, in 
the dimer state for the square substrate [Fig.~1(b)].
As illustrated in Fig.~3(a) at low temperatures $T/T_{m}^{0}<0.25$ 
(where $T_{m}^{0}$ is the melting temperature at zero substrate
strength), both orientational
and translational order are present and the system is frozen. 
We label this
phase the ordered solid. 
In Fig.~3(b), at $T/T_{m}^{0}=1.5$, 
the dimers begin to rotate within the minima; however, 
diffusion of individual colloids 
throughout 
the sample does not occur so the system is still
frozen with the dimer orientational order lost.  We label this phase the
partially ordered solid.  
For a higher temperature $T/T_{m}^{0}=4.0$
[Fig.~3(c)], the system enters a modulated liquid phase.  Here the 
colloids began to diffuse throughout the system.
We have observed this same type of multi-stage melting for the other CMC's in 
Figs.~1 and 2. 

In Fig.~4 we show the phase diagram of the temperature $T/T_{m}^{0}$ 
vs substrate 
strength $A$ for the case of a square substrate and $N_{c}=2N_{m}$.
The melting line is determined from the onset of diffusion, and
the solid to disordered solid 
transition line is determined from the correlation
between the dimers as well as the onset of dimer rotation,
measured using the average particle displacements.
For zero substrate $A=0$, the colloids
form a triangular 
lattice and the clean melting temperature $T^{0}_{m}$ is
determined from the onset of particle diffusion.  
For increasing $A$ the melting temperature monotonically increases. The 
ordered solid melts directly to the liquid for $A < 2.0$. 
For $A > 2.0$ an intermediate rotational melting transition
between the ordered solid and partially ordered solid occurs. The rotational
melting temperature is always less than the clean melting line, since
both these melting temperatures are determined by the elastic properties
of the colloids, which is maximum 
for a triangular lattice in the 
zero substrate
limit. The ordered solid is not triangular and hence should have less 
elasticity.  In addition, the 
phase
diagram shows that for increasing $A$ the transition line from the
ordered solid to partially ordered solid 
decreases in temperature.  This occurs since, as $A$ is increased, the
dimer length decreases, the dimers become effectively further
apart, and the dimer-dimer interaction strength goes down. 
This result also implies that at fixed 

\begin{figure}
\center{
\epsfxsize=3.5in
\epsfbox{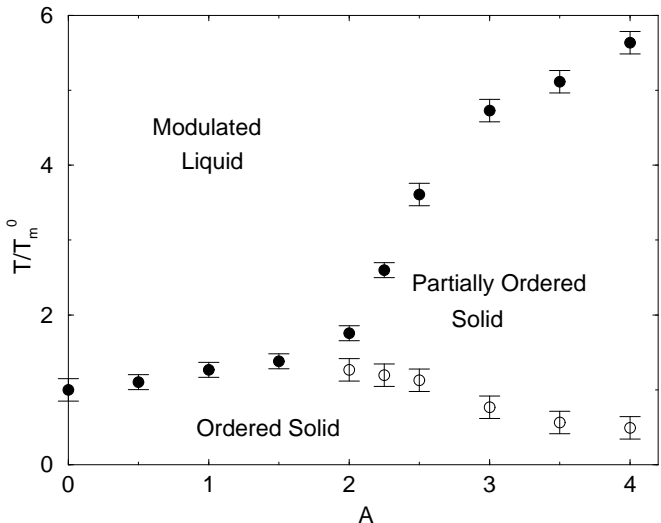}}
\caption{  
Phase diagram of temperature $T/T^{0}_{m}$ 
vs substrate strength, for the square substrate with two colloids per
potential minima. 
$T_{m}^{0}$ is the melting temperature for a sample with no substrate.  
Black circles: The melting line 
as determined from the onset of diffusion.  Open circles: The 
transition to the partially ordered or rotating dimer phase as determined
from the orientational correlation of the dimers.  
}
\end{figure}

\hspace{-13pt}
temperature it is possible to 
observe a transition 
from
the ordered to partially ordered solid by increasing the strength of the 
substrate. This is similar to the reentrant melting observed for
increasing substrate strength for colloids in 1D periodic substrates 
\cite{Leiderer,Frey}.  
We find qualitatively  
similar phase diagrams for square and triangular substrates for 
both the dimer and trimer states. For the trimers the onset 
of rotational melting occurs at weaker substrate strengths
than for the dimers, due to the decreased anisotropy
of the trimer state compared to the dimers. 
In addition for stronger screening the 
transition from the ordered solid to the partially ordered solid occurs
for both lower temperature and substrate strength. 
An interesting issue is the nature of the melting transitions, 
which is
beyond the scope of this work. 
   
To summarize, we have demonstrated that a rich variety of novel colloidal
crystalline states can be achieved with square and triangular two dimensional
substrates. We find that when there are more colloids than substrate potential 
minima, certain colloidal molecular crystals appear in which each minima 
captures an equal number of colloids.
The colloids in these traps can act as a single particle with a 
rotational degree of freedom, such as a dimer or trimer.  
We also show that CMC states exhibit a multi-stage melting 
for sufficiently strong substrates, where the 
orientational order of the colloidal molecule states is lost first 
followed by the translational order.
We  map out the phase diagram and show that the solid to 
disordered solid transition occurs for temperatures lower than the
clean melting temperature. 
It is also possible to have a transition at fixed temperature
for increasing substrate strength 
from a ordered solid to a partially ordered solid, similar
to the reentrant melting observed for colloids on 1D periodic substrates. 
Since the colloids within a minima can act as a single particle with 
a rotational degree of freedom, 
our results also suggest that certain canonical statistical mechanics models, 
such as Ising, XY, Potts, and frustrated models, may be 
realized with colloids on two
dimensional periodic substrates. 
The states predicted here should be observable for colloids interacting 
with crossed laser arrays or optical tweezer arrays, dusty plasmas in
2D with periodic potentials, and vortices in superconductors with periodic
substrates. 

We thank, C. Bechinger, 
D.G.~Grier, P.~Korda, S. Ling, A. Persinidis, and G. Spalding 
for useful discussions. 
This work was supported by the US Department of Energy
under contract W-7405-ENG-36.

\end{document}